\documentstyle[12pt,aasms4]{article}
\lefthead{JONES AND WEHRLE}
\righthead{VLBA IMAGING OF NGC 4261} 
\begin{document}
\title{VLBA Imaging of NGC 4261: Symmetric Parsec-scale Jets \\
and the Inner Accretion Region} 

\author{Dayton L.~Jones} 
\affil{Jet Propulsion Laboratory, California Institute of Technology} 
\authoraddr{Mail Code 238-332, 4800 Oak Grove Drive, Pasadena, CA 91109;
dj@bllac.jpl.nasa.gov} 
\and 
\author{Ann E.~Wehrle}
\affil{Infrared Processing and Analysis Center, Jet Propulsion Laboratory,\\
California Institute of Technology} 
\authoraddr{Mail Code 100-22, California Institute of Technology, 
770 S.~Wilson Ave., Pasadena, CA 91125;
aew@ipac.caltech.edu}

\begin{abstract}  
We observed the nuclear region of NGC 4261 (3C270) with VLBI to 
determine the morphology of the central radio source on parsec scales,
and in particular to see if the inner radio axis remained in the same
direction as the kpc-scale jets or whether it was aligned with the
apparent rotation axis of the nuclear disk imaged by HST.  The position
angle of the radio axis in our VLBA images agrees, within the errors,
with the position angle of the VLA-scale jet.  Thus, there is no 
evidence for precession of the jets on time scales shorter than the
material propagation time from the nucleus to the diffuse radio lobes.
Our dual frequency observations also reveal basically symmetric   
radio structures at both 1.6 and 8.4 GHz.  Analysis of these images  
shows that most of the central 10 pc of this source
is not significantly affected by free-free absorption, even though
HST images of the the nucleus of the galaxy show it to contain a nearly 
edge-on disk of gas and dust on larger scales.  The lack of detectable 
absorption over most of the central 10 pc implies that the density of 
ionized gas in this region is less than $\sim 10^{3} \ {\rm cm}^{-3}$,
assuming a temperature of $\sim 10^{4}$K.  Our highest angular 
resolution images show a very narrow absorption feature just east of the
radio core, suggesting that there may be a small, dense inner accretion
disk whose width is less than 0.1 pc.   
If the inclination of this inner disk is close to that of the 
larger-scale HST disk it becomes optically thin to 8.4 GHz radiation
at a deprojected radius of about 0.8 pc.
The brightness of the pc-scale jets falls off very rapidly on both sides
of the core, suggesting that the jets are rapidly expanding during
the first several pc of their travel.  The rate of jet expansion must
slow when the internal pressure falls below that of the external 
medium.  We suggest that this occurs between about 10 and 200 pc 
from the core because the rate of decrease in radio brightness is
far slower $>$ 200 pc from the core than it is within 10 pc of the core.  
It appears that there is a small dense inner disk centered on the radio
core (the base of the jets; $< 1$ pc), a low density, presumably hot 
``bubble" filling most of the the inner several pc of the nucleus 
(within which the radio jets expand rapidly; $\sim 10$ pc), and a 
surrounding cool, higher density region (of which the HST absorption 
absorption disk is part; $> 10$ pc) within which the transverse  
expansion of the radio jets, as implied by the rate of decrease in
jet brightness, is nearly halted.  

\end{abstract}


\section{Introduction}

The radio source 3C270 (PKS 1216+06) associated with the E2 galaxy 
NGC 4261 is composed of two symmetric lobes of extended emission
on opposite sides of the galaxy, connected by symmetric kpc-scale
jets to a compact radio source coincident with the optical nucleus
of the galaxy. 
The compact radio core is a relatively weak VLBI source 
(180 mJy at 1.6 GHz; Jones, Sramek, and Terzian\markcite{1} 1981). 
The radio jets extend out approximately $\pm 4$ arcmin, have 
opening angles of less than $5^{\circ}$, and 
are co-aligned to within $1^{\circ}$ along position angle $88^{\circ}$
(Birkinshaw and Davies\markcite{2} 1985).  The ratio of jet/counterjet
surface brightness within a few arcseconds of  
the central compact source is only about 2:1, 
indicating that relativistic beaming effects are not strong on kpc scales. 
The minor axis position angle of the galaxy remains $69^{\circ} \pm 2$
over a wide range of radii 
(van den Bosch, et al.\markcite{32} 1994; Ferrarese, et al.\markcite{30} 
1996), $19^{\circ}$ from the radio axis, but the stellar rotation axis is 
along position angle $153^{\circ} \pm 4$ (Davies and Birkinshaw\markcite{4} 
1986) only $6^{\circ}$ from the projected major axis of the galaxy.  
Evidently NGC~4261 is a nearly prolate galaxy whose
stellar rotation axis has no relationship with the direction of the radio 
jets.  

M\"{o}llenhoff and Bender (\markcite{7} 1987) discovered a small dust 
lane in the center of NGC~4261 which is oriented perpendicular to
the radio jets.  More recent HST observations have revealed that  
the optical nucleus is surrounded by a disk of gas and dust 
1.7 arcsec in diameter whose projected 
rotation axis is aligned within several degrees of the  
radio jets (Jaffe, et al.\markcite{8} 1993; 
1996; Ferrarese, et al.\markcite{30} 1996).  
At a distance of 41 Mpc (Faber et al.\markcite{9} 1989)
the HST disk diameter is 340 pc.  Some authors give distance 
estimates up to three times smaller than this, which of course 
would reduce the calculated physical size of the HST disk by the same 
factor.  The rotation axis of the HST disk is inclined $69^{\circ}$
from our line of sight based on HST FOS spectral data,  
or $64^{\circ}$ based on isophote fitting
(Ferrarese, et al.\markcite{30} 1996).  These values agree 
within their errors.  This suggests  
that the radio axis may be at a similar angle from our line of sight. 

The apparent inclination of the HST absorption 
disk and the larger-scale (M\"{o}llenhoff and Bender dust lane) 
disk differ by about $10^{\circ}$. 
It is likely that the HST disk and the 
larger scale dust lane are both part of a single warped disk 
structure (Mahabal et al.\markcite{11} 1996). 
Since the minor axis of the HST disk is at a position angle of
$63^{\circ} \pm 2^{\circ}$ (Ferrarese et al.\markcite{30} 1996), 
which is not parallel to the position angle of the radio jets 
($88^{\circ} \pm 1^{\circ}$; Birkinshaw and Davies\markcite{2}  
1985), the presumed warp continues into the region 
close to the central black hole.  
Jaffe et al.\markcite{10} (1996)
point out that the apparent center of the HST disk is displaced from
the optical center of the galaxy (based on isophote fitting) by at 
least 5 pc (see also Ferrarese et al.\markcite{30} 1996).

Neutral hydrogen and CO have been detected in absorption against
the radio core by Jaffe and McNamara\markcite{12} (1994).  Their
measurements indicate that the total mass of gas in the HST disk
is $\sim 10^{5}$ M$_{\odot}$, which is sufficient to power the
radio source for $\sim 10^{8}$ years.  All of the above evidence
suggests that the radio source is powered by material accreting
in from the large-scale dust disk through the HST dust disk and
eventually onto the (unseen) inner accretion disk where the radio
jet formation takes place within a few
Schwartzschild radii of a central massive black hole.  The 
accreting material probably came from a merger between NGC~4261 
and a smaller gas-rich galaxy.  This would explain the apparently
unrelated dynamics between the gas and dust in the nuclear disks
and the over-all stellar dynamics of the galaxy. 

We observed the nuclear region of NGC~4261 with VLBI to determine
the morphology of the central radio source on parsec scales, and 
in particular to see if the inner radio axis remained in the same 
direction as the kpc-scale jets (position angle = 88$^{\circ}$) 
or whether it was aligned with the apparent
rotation axis of the HST disk. 

\section{Observations and Results}

We observed NGC~4261 for 8.5 hours on 1 April 1995 using all ten 
antennas of the NRAO Very Long Baseline Array.  Individual scans
of typically 26 minutes duration were alternated between 1.6 GHz
and 8.4 GHz.  We recorded a bandwidth of 64 MHz with 
single polarization, and the data were cross-correlated on the
VLBA correlator.

After cross-correlation the data were read into 
AIPS\footnote{The Astronomical Image Processing System
was developed by the National Radio Astronomy Observatory, which
is operated by Associated Universities, Inc., under a cooperative
agreement with the National Science Foundation.}, where standard
programs were used for editing, amplitude calibration, and fringe
fitting.  Prior to fringe fitting the strong compact radio source 
1308+326 was used to derive corrections for phase slopes across the 
frequency channels at both 1.6 and 8.4 GHz.  After fringe fitting
the data were averaged over frequency and exported to the Caltech 
program Difmap (Shepherd, Pearson, and Taylor\markcite{13} 1994) 
for additional editing, self-calibration, imaging, and deconvolution. 
Only marginal detections were obtained on the longest VLBA baselines
(to the antennas at Saint Croix and Mauna Kea) at 1.6 GHz, but 
good fringes were found to all ten antennas at 8.4 GHz.  

Our final VLBA images are shown in figures 1-5.  Figures 1 and 2 
are the highest dynamic range images at 1.6 and 8.4 GHz, respectively. 
At both frequencies the source appears two-sided, although the 
extension to the west (the direction of the slightly brighter VLA 
jet) falls off more slowly in brightness
than the extension to the east. 
In neither image does the source appear to have a one-sided
``core-jet" morphology.
At a distance of 41 Mpc, 1 milliarcsecond (mas) corresponds to 
0.20 pc.  Consequently, the lengths of the visible radio jets 
(measured from the central brightness peak) are
approximately 12 pc at 1.6 GHz and 4 pc at 8.4 GHz.

\placefigure{fig1}

\placefigure{fig2}

Figure 3 is our highest resolution image at 8.4 GHz made with 
full weighting of the longest baselines.  Because the longest 
baselines have the lowest average signal-to-noise ratios,
the noise level in figure 3 is nearly a factor of two greater 
than in figure 2.  However, figure 3 does reveal a new feature
which is not visible in lower resolution images: the very sharp
eastern edge of the brightest peak.  Our gray-scale image 
(figure 4) shows this 
to be a narrow gap in emission between the peak and
the first obvious feature of the east-pointing jet.  No such
gap is seen on the western side of the source.  Thus, the 
radio structure in figures 3 and 4 appears to be much less
symmetric than that in figures 1 and 2.  
This is the expected signature of a small, dense inner accretion 
disk, whose apparent thickness is less than 0.5 mas or 0.1 pc.
If real, this inner disk becomes optically thin at 8.4 GHz at a
projected radius of about 1.5 mas (0.3 pc); the deprojected
radius is smaller than 1 pc unless the inclination angle 
of the inner disk is larger than $72^{\circ}$ (recall that
the inclination angle of the much larger HST disk is  
$\approx 67^{\circ}$).

\placefigure{fig3} 

\placefigure{fig4}

If our interpretation of this gap as being the result of free-free
absorption by an inner accration disk is correct, it is perhaps
surprising that an even larger absorption feature is not seen in
our 1.6 GHz images.  To investigate this, we simulated the 
one-dimensional brightness distribution of an intrinsically
symmetric source with a narrow gap on one side and convolved 
this profile with Gaussian beams of different widths.  We found
that it is possible for the simulated absorption feature
to be easily visible with the VLBA 8.4-GHz beam but invisible 
with the 1.6 GHz beam, which is more than five times as wide. 
In fact, the actual beam used at 1.6 GHz is nearly ten times as
wide as the full-resolution 8.4 GHz beam because we downweighted
the low SNR data on the longest baselines at 1.6 GHz to improve
the dynamic range. 

Figure 5 is an image at 8.4 GHz which was made by applying a 
Gaussian taper to the visibility data to match the angular 
resolution available at 1.6 GHz.  The same field of view and
restoring beam has been used for figures 1 and 5 to allow 
easy comparison.  

\placefigure{fig5}

Although the degree of symmetry depends on both frequency 
and angular resolution, the source appears basically two-sided 
in all of the above figures.  The more symmetric appearance 
at 1.6 GHz is partially caused by the greater extent of the 
jets at 1.6 GHz due to spectral index effects, and also 
suggests that the cause of the ``gap" in figures
3 and 4 is confined to an angular scale much smaller than the
1.6 GHz or tapered 8.4 GHz beams.

Attempts were made to detect more distant emission at both
1.6 and 8.4 GHz using larger image sizes and tapering the
visibility data to favor short baselines.  In no case did
we find any emission significantly more extended than that
shown in figures 1 and 5, although the range of VLBA baselines
would allow more extended radio structure to be detected.  We
conclude that the brightness of the two parsec-scale radio
jets does drop below our noise level very rapidly at both
frequencies, and that this is not an artifact of the imaging
procedure.  

In referring to the images in figures 1-5 as symmetric or 
two-sided, we are implicitly assuming that the central brightness
peak is the ``core" of the source (the base of the radio jets).
Is there any evidence to support this assumption?  
We have compared the images in figures 1 and 5 to determine spectral 
index distributions for a range of position offsets between the
two images.  Figure 6 shows the spectral index distribution along
the east-west axis of the source when the central peaks at both  
frequencies are aligned.  Not surprisingly, the spectral index
distribution is also quite symmetric with an inverted spectrum
near the center and increasingly steep spectra with increasing
distance from the center.  It is possible to offset the 8.4 GHz
peak up to 10-12 mas from the 1.6 GHz peak without making the
spectral index $\alpha > +2.5$ (using ${S_{\nu}} \propto 
{{\nu}^{\alpha}}$), but in this case the spectral index more
than $\approx 30$ mas from the center becomes extremely steep
($\alpha < -2$).  We therefore favor a registration in which
the brightest peaks in figures 1 and 5 are nearly co-aligned,
giving a spectral index distribution close to that shown in
figure 6.  

\placefigure{fig6}

The inverted spectrum at the center of the source could be 
caused by synchrotron self-absorption or by free-free absorption.
In the case of free-free absorption the ionized gas responsible  
for the inverted spectrum at the center
of the source would have to cover only the inner 0.2-0.3
pc and/or have a filling factor substantially less
than unity to account for the fact that $\alpha < 1$.  Free-free
absorption by thermal gas uniformly covering a source gives an
inverted spectrum with $\alpha \approx 2$.
Synchrotron self-absorption is seen in the cores of many 
compact extragalactic radio sources, and by analogy we might
expect to see this effect in the core of NGC 4261.  Our brightness
temperatures (see below) are lower limits, and therefore they do not 
rule out synchrotron self-absorption, which can produce inverted spectra 
with $\alpha \le 2.5$ depending on the turn-over frequency
and homogeneity of the source.  We can not exclude a combination
of both effects in the center of NGC 4261.  
In both morphology and spectral index distribution NGC~4261 is
very similar to the parsec-scale source in Hydra A 
(Taylor\markcite{14} 1996).

The brightness of both the east and west jets drops 
off rapidly at both frequencies.  To quantify this, we found
that three of the four brightness profiles could be reasonably well 
fit with a power law of the form ${\rm (surface\ brightness)} \propto 
{{\rm (distance\ from\ peak)}^{x}}$.  At 1.6 GHz the exponent
$x$ is -1.9 for the jet extending to the west and -2.0 for
the jet extending to the east.  The linear brightness profile
fits for 1.6 GHz are shown in figure 7.

\placefigure{fig7}

Although the western jet fades slightly  
more slowly with distance from the core, these values
are very similar.  At 8.4 GHz the corresponding exponents
are $x$ = -1.2 for the western jet and -2.0 for the eastern
jet.  Again, the eastern jet fades more rapidly.  The eastern
jet at 8.4 GHz is not very well fit by a single exponential law,
although the western jet is. 
The 8.4 GHz brightness profile
fits are shown in figure 8.

\placefigure{fig8}

In all cases we used measurements of the brightness per unit
length (Jy/mas) starting near 50\% of the peak value
and extending out in steps of 0.7-0.8 times the half power
beamwidth (6.5 mas steps at 1.6 GHz and 1.2 mas steps at 8.4 GHz).
This represents a compromise between having truly independent 
measurements and having an adequate number of 
points for the least-squares fit. 

The total flux density of the VLBI structure shown in figures
1 and 3 is 0.20 Jy at 1.6 GHz and 0.34 Jy at 8.4 GHz.  The 
peak (core) surface brightness is 0.10 Jy/beam at both 1.6 GHz
and 8.4 GHz.  This corresponds to a brightness
temperature of $3.3 \times 10^{8}$K at 1.6 GHz and 
$1.2 \times 10^{9}$K at 8.4 GHz; these values are lower 
limits to the true brightness temperature of the core, which
is angularly unresolved at both frequencies.  

The VLBI position angles are $86.2^{\circ} \pm 1.5^{\circ}$ at 
1.6 GHz and $85.6^{\circ} \pm 1.5^{\circ}$ at 8.4 GHz.  At both 
frequencies the VLBI position angles are consistent (within the 
combined errors) with the $88^{\circ} \pm 1^{\circ}$ position 
angle of the VLA jets, although the 8.4 GHz value may suggest a 
small curvature of the jets very close to the core.  If real, this 
curvature makes the inner jet orientation slightly closer to the 
minor axis of the HST disk ($73^{\circ} \pm 2^{\circ}$; Ferrarese 
et al.\markcite{30} 1996).  Higher frequency VLBI observations 
will be needed to determine the reality of any jet curvature in 
the central few pc of this source.  

\section{Discussion}

It is clear from VLA images that both the east and west jet in
NGC~4261 extend far beyond the scale of our VLBA images.  
Therefore, the disappearance of both jets within a few tens
of mas of the core is not caused by their disruption or 
``smothering" by a dense interstellar medium.  It is 
possible that the jets are fading due to expansion  
close to the core.  This implies that the external pressure
is less than the internal jet pressure.  At some point the
internal pressure in the expanding jets will become lower 
than the external pressure (which should fall more slowly 
than $d^{-2}$ at large distances), causing the opening angle 
of the jets to be reduced and possibly creating shocks which
could reaccelerate the relativistic electrons.  The result
will be a much slower decrease in jet brightness on kpc scales.
Our angular resolution is insufficient to measure the opening
angle of the pc-scale jets, but from figure 2 we can get an
upper limit of approximately $10^{\circ}$ for the east jet,
compared with less than $5^{\circ}$ on kpc scales.

We can set an upper limit on the free electron density in 
the inner $\sim 10$ pc of NGC~4261 from the apparent lack of 
free-free absorption of either jet at 1.6 GHz.  If absorption 
by a 10-pc-scale disk 
or torus of ionized gas were significant, we would expect to see
a much more asymmetric radio morphology in figures 1 and 2 
({\it cf.} 3C84; 
Vermeulen, Readhead, and Backer\markcite{15} 1994; Walker,
Romney, and Benson\markcite{16} 1994).  
Since both east and west jets 
in NGC~4261 are visible out to $\sim 10$ pc with similar
brightness, the optical depth from 
free-free absorption must be much less than unity.  Assuming
a gas temperature of $\sim 10^{4}$K and a path length of 10 pc
gives an electron density $< 10^{3}$ cm$^{-3}$.  
Note however that our limit applies to the inner few pc of 
NGC~4261 as a whole; the apparent absorption on a sub-pc scale
in figures 3 and 4 would be undetectable at lower angular 
resolutions.  Thus, we are not ruling out 
free-free absorption by a high electron density gas within the 
central pc.  If the path length through the inner disk suggested
by figures 3 and 4 is assumed to be $\sim 0.1$ pc, the resulting
lower limit for $n_{\rm e}$ is $\sim 10^{5}$ cm$^{-3}$.

For NGC~4261 the total mass of ionized gas implied in the inner
10 pc, neglecting the possible sub-pc inner disk, is 
$< 10^{5}$ M$_{\odot}$.  It appears
that we have a low density, presumably hot, inner ``bubble"
filling the inner several pc of the nucleus (within which the
radio jets expand rapidly) surrounded by a cool, higher density
region (of which the HST absorption disk is part) within 
which the transverse expansion of
the radio jets is nearly halted.     
Higher electron densities have been estimated from optical 
observations (Jaffe, et al.\markcite{10} 1996; Ferrarese, 
et al.\markcite{30} 1996), but the angular resolution of these
measurements is no better than 0.1 arcsec and consequently they
may be partially sampling a higher density region surrounding the 
region in which we see the VLBI radio jets.
Another possibility is that much of the optical emission 
comes from within the central pc, in which case a small region 
of higher density gas could dominate the optical measurements.  

As Bridle and Perley (\markcite{17} 1984) point out, the central
brightness of an expanding synchrotron jet decreases as a 
different power of distance (or radius) $d$ depending on whether 
the jet is dominated by a parallel or transverse magnetic field,  
or if equipartition is assumed.  Our VLBI images do  
not resolve the jets in the transverse direction, so we measure
the brightness per unit length rather than the surface brightness. 
For an optically thin jet with $\alpha = -0.65$
the brightness per unit length falls off as $d^{-4.2}$ for a 
parallel magnetic field, $d^{-2.5}$ for a transverse magnetic
field, and $d^{-3.1}$ if equipartition is assumed.  All of 
these exponents are more negative than the observed fall-off 
of brightness per unit length in the parsec-scale jets of
NGC 4261.  This is a common result for many radio jets (e.g.,  
Perley, Bridle, and Willis\markcite{33} 1984), and it implies
than magnetic field amplification and/or particle reacceleration
(or a dramatic reduction in jet velocity) must take place as 
the jet expands.  Based on figure 5, the  
jets in NGC 4261 appear to be optically thin only at distances
greater than 10-15 mas from the core.  This complicates the 
interpretation of the brightness profiles at 8.4 GHz, but at
1.6 GHz the jet profiles with their $d^{-2}$ fall-off extend 
well beyond this region.  

Taylor (\markcite{14}1996) presents VLBA images and spectral
index maps of the central radio source in Hydra A which look
remarkably similar to those presented in this paper.  Taylor
finds evidence for significant ($\tau > 1$) free-free 
absorption of the radio core in Hydra A at 1.3 GHz and 
deduces a value of $r \ {n_{e}}^{2}$ which is equal to the
upper limit we find for NGC~4261.
At our lowest frequency of 1.6 GHz the free-free absorption
deduced by Taylor would have an optical depth $< 1$.  If the
thermal gas density in the nucleus of NGC~4261 is somewhat 
smaller than in Hydra A, the effects of free-free absorption
would be quite small at our observing frequencies.  The neutral
hydrogen column density in front of the radio core in NGC~4261  
is approximately 20 times smaller than in Hydra A (Jaffe and
McNamara\markcite{12} 1994; Taylor\markcite{14} 1996). 

Another method for estimating hydrogen column density is
through X-ray absorption.  
NGC~4261 contains an X-ray source whose angular extent is 
similar to that of the optical galaxy (Fabbiano, Kim, and
Trinchieri\markcite{19} 1992; Worrall and Birkinshaw\markcite{40} 
1996), as well as fainter emission on larger scales (Worrall
and Birkinshaw\markcite{21} 1994; Davis, et al.\markcite{23} 1995)
and a strong component which is angularly unresolved by ROSAT 
(Worrall and Birkinshaw\markcite{40} 1994; 1996).
Birkinshaw and Worrall (\markcite{24}1996) concluded that the 
extended X-ray gas confines the kpc-scale radio jets, and that the 
jets terminate (flair into the two diffuse lobes) at the point where
the external pressure falls below the internal pressure in the
jets.  This implies that the flow in the jets is subsonic.
Einstein IPC spectra can be 
fit with a power law (Kim, Fabbiano, and Trinchieri\markcite{20}
1992), but a thermal plus power law spectrum provides the best
fit to ROSAT data (Worrall and Birkinshaw\markcite{21} 1994).  
In the thermal plus power law model the derived neutral hydrogen
column density in NGC~4261 is quite small 
(${{\rm N}_{H}} < {4 \times 10^{20}}$ atoms cm$^{-2}$), and the
cooling time for gas within the core radius is estimated to be
$3.4 \times 10^{9}$ years.  
The low hydrogen column density derived from the X-ray data are
difficult to reconcile with the higher HI densities derived from  
the radio absorption measurements of Jaffe and McNamara\markcite{12}
(1994).  Worrall and Birkinshaw\markcite{21} (1994) discuss several
possible reasons for this difference.  
One explanation is that the radio emission 
comes to us directly from the core and inner regions of the jets,
and consequently at least partly through the suggested dense 
inner accretion disk, 
while the X-ray emission (the nonthermal power law component) 
is initially directly in the same general direction as the radio
jets and is scattered into our line of sight by plasma within 
the jet or surrounding the inner nuclear region.  If this 
proposed X-ray scattering occurs far enough from the radio core,
the gas density along the mean X-ray line of sight could be 
substantially smaller than along the mean radio line of sight.  
This model predicts that future very high angular resolution X-ray
images of the NGC~4261 nucleus will find a position offset between
the radio and X-ray peaks, and that this offset will be approximately
along the direction of the radio jets.  
An alternative possibility is that the X-ray emission comes from an
angularly larger region than the bright central radio source; this
would have the same effect of allowing most of the X-ray emission 
to avoid the presumed inner accretion disk.  

The position angle of the radio axis in our VLBA images agrees,
within the combined errors, with the position angle of the VLA-scale 
jets.  Thus there is no evidence for precession of the jets on time  
scales shorter than the material propagation time from the 
nucleus to the diffuse radio lobes. 
This long-term stability implies that the central compact 
object has a very large angular momentum. 

If the central black hole in NGC~4261 is $10^{8}-10^{9}$ M$_{\odot}$
it is unlikely that it would have any detectable orbital motion
about a presumably less massive secondary nucleus from a merger 
(which could be the cause
of the positional offset between the apparent center of the HST
dust disk and the radio core).  In this case the secondary nucleus
(and perhaps parts of the nuclear gas disk) would show almost all of the
orbital motion.  This could also explain the small but significant
difference between the radio jet position angle and the position
angle of the HST disk rotation axis.    

\section{Conclusions}

We have found that the pc-scale radio source in the nucleus of
NGC~4261 is unusually symmetric, and aligned along the same 
position angle as the larger-scale radio structure imaged with
the VLA.  The morphology and spectral index distribution of the
pc-scale source indicates that free-free absorption is not
significant within several pc of the radio core, except possibly
within the central pc.  The 
brightness of both radio jets decreases very rapidly in both 
directions from the core, which we interpret as evidence for
a large initial opening angle (rapid expansion).  
At some distance between about 50 and 500 pc the jet opening
angles decrease to the $< 5^{\circ}$ value seen on kpc scales.
This should occur when the internal jet pressure falls
below that of the external medium.    

\acknowledgments

We thank David Meier for several informative discussions about
jet collimation, D.~Worrall for information on the large scale
X-ray morphology, and the anonymous referee for several corrections
and suggestions which significantly improved this paper.  
The Very Long Baseline Array is part of the National Radio
Astronomy Observatory, which is a facility of the National
Science Foundation operated by Associated Universities, Inc.,
under a cooperative agreement with the NSF.  
A.W.~gratefully acknowledges support from the NASA Long Term 
Space Astrophysics Program.
This research was carried out at the Jet Propulsion
Laboratory, California Institute of Technology, under contract with
the National Aeronautics and Space Administration.  

\clearpage

\figcaption{VLBA image of NGC~4261 at 1.632 GHz.  The contour levels
are -0.5, -0.25, 0.25, 0.5, 1, 2, 4, 8, 16, 32, 50, 70, and 95\% of
the peak surface brightness (98 mJy/beam).  The restoring beam is
$14.70 \times 9.11$ mas with the major axis along position angle 
-10.5$^{\circ}$.
\label{fig1}} 

\null

\figcaption{VLBA image of NGC~4261 at 8.387 GHz.
The contour levels are -0.25, 0.25, 0.5, 1, 2, 5, 8, 16, 32, 50, 70,
and 95\% of the peak surface brightness (149 mJy/beam).  The restoring
beam is $2.69 \times 1.56$ mas with the major axis along position angle
-2.2$^{\circ}$.
\label{fig2}}

\null

\figcaption{Full resolution VLBA image of NGC~4261 at 8.387 GHz. 
The contour levels are -0.5, 0.5, 1, 2, 4, 8, 16, 32, 70, 70, and 95\%
of the peak surface brightness (101 mJy/beam).  The restoring beam is
$1.84 \times 0.80$ mas with the major axis along position angle 
-1.1$^{\circ}$. 
\label{fig3}}

\null

\figcaption{Gray scale representation of the full-resolution 8.387 GHz 
image, showing the narrow gap in emission just east of the brightest
peak (core). 
\label{fig4}}
 
\null

\figcaption{Low resolution VLBA image of NGC~4261 at 8.387 GHz.  The contour 
levels are -0.12, -0.06, 0.06, 0.12, 0.25, 0.5, 1, 2, 4, 8, 16, 32, 50, 70,
and 95\% of the peak surface brightness (288 mJy/beam).  For comparison
with figure 1 the same field of view and restoring beam ($14.70 \times 9.11$
mas, position angle = -10.5$^{\circ}$) has been used. 
\label{fig5}}  

\null

\figcaption{The spectral index $\alpha$ between 1.6 GHz and 8.4 GHz as a
function of distance along the radio axis (PA = $88^{\circ}$).  These
values were determined by aligning the brightest peak at both 
frequencies (see figures 1 and 4).  
The small peak about 32 mas east (right) of the core is caused by the 
knot of emission visible in the eastern jet at 8.4 GHz (see figures
2 and 3).  
\label{fig6}}

\null

\figcaption{Profile of brightness per unit length at 1.6 GHz along
position angle $87^{\circ}$.
{\bf a)} \ East-pointing jet, {\bf b)} \ West-pointing jet.
The slope of the least-squares fit line is -2.0 for the east
jet and -1.9 for the west jet.  The distance between measurements
is 6.5 mas except for the innermost point of the east jet.  The
projected restoring beam is a Gaussian with 9.23 mas full width
at half maximum.
\label{fig7}}

\null

\figcaption{Profile of brightness per unit length at 8.4 GHz along
position angle $87^{\circ}$.
{\bf a)} \ East-pointing jet, {\bf b)} \ West-pointing jet.
The slope of the least-squares fit line is -2.0 for the east
jet and -1.2 for the west jet.  Note that at this frequency the
east jet is not well fit by a single power law.  
The distance between measurements is 1.2 mas and the projected
restoring beam is a Gaussian with 1.56 mas full width at half
maximum.  
\label{fig8}} 

\null


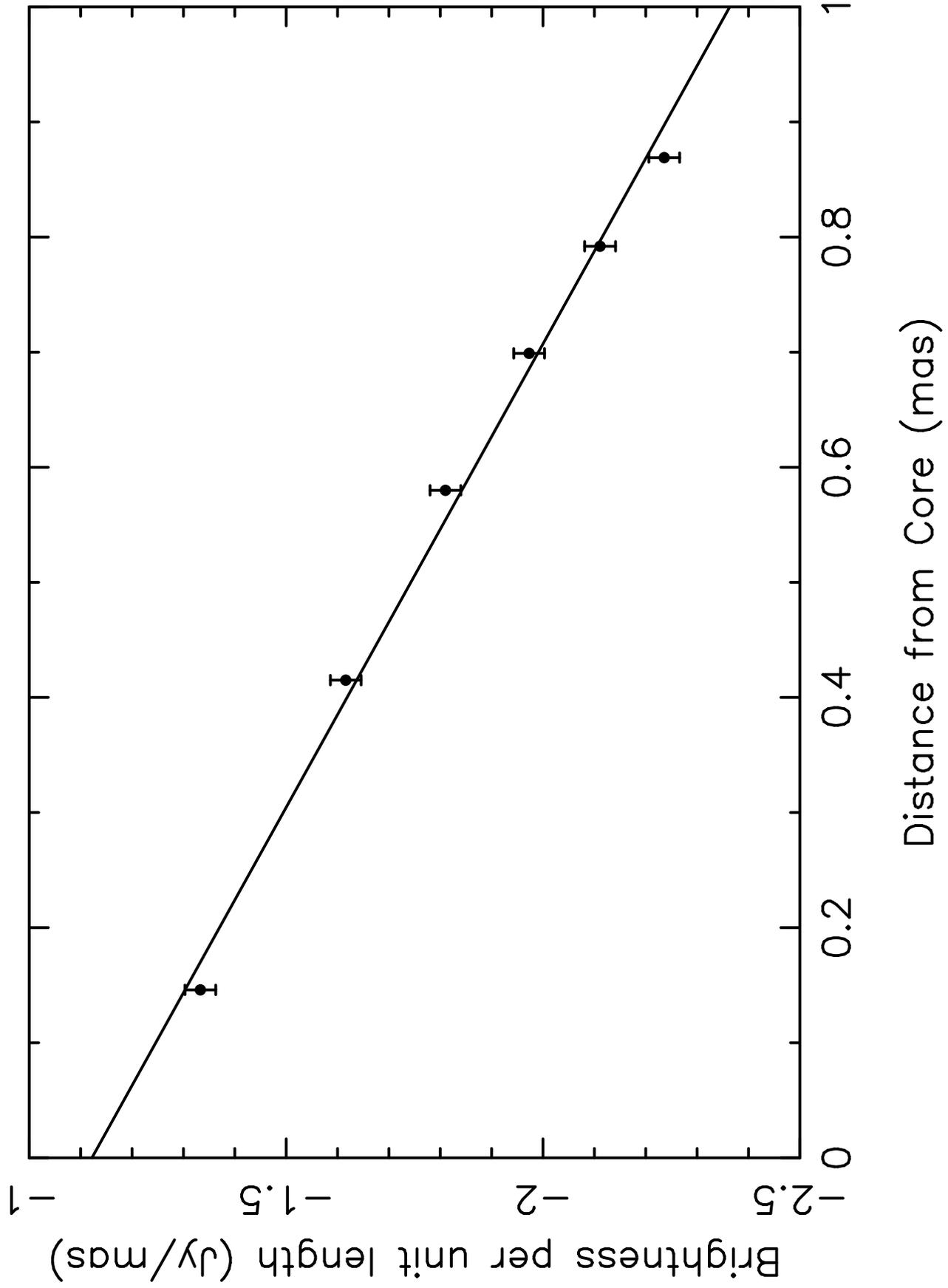
\end{document}